\documentclass[twocolumn,prb,showpacs]{revtex4}
\usepackage{graphicx}
\usepackage{dcolumn}
\usepackage{bm}
\usepackage{amsmath}

\begin{document}

\preprint{}
\title{ Transverse Josephson effect due to spin-orbit coupling: Generation of transverse current without time-reversal symmetry breaking}
\author{Takehito Yokoyama}
\affiliation{Department of Physics, Tokyo Institute of Technology, Tokyo 152-8551,
Japan 
}
\date{\today}

\begin{abstract}
We investigate transverse Josephson current in superconductor/normal metal/superconductor junctions where the normal metal has Rashba type spin-orbit coupling.  It is shown that transverse current arises from the spin-orbit coupling in the normal metal. 
This effect is specific to superconducting current and the transverse current vanishes in the normal state.  
In addition, this transverse Josephson effect is purely stationary and applied magnetic field is unnecessary to realize this effect, in contrast to the Hall effect in the normal state. We also discuss physical interpretation of this effect, comparing with the spin Hall effect. 

\end{abstract}

\maketitle


When magnetic field is applied to electrons,  the Lorentz force acts on the electrons and a voltage drop appears in the direction perpendicular to the applied current. This is the celebrated (classical) Hall effect. 
About 100 years later, the quantum Hall effect has been discovered\cite{Klitzing,Tsui,Prange}: When a strong magnetic field is applied to 2D electron gas perpendicularly, the longitudinal resistance vanishes while the Hall conductance is quantized to a rational multiple of $e^2/h$. 
The classical and quantum Hall effects arise by applying magnetic field. Therefore, the Hall effect occurs together with the time-reversal symmetry breaking.  
Several authors have proposed the models which exhibit a nonzero quantization of the Hall conductance by a spatially inhomogeneous magnetic flux with zero average and hence without Landau levels\cite{Haldane,Tang,Neupert,Sun}.
In these models, time-reversal symmetry is also broken due to the inhomogeneous magnetic flux.
On the other hand, there have been a few efforts to realize the Hall effect without any magnetic field (flux). 
It has been shown that a circularly polarized light radiation can induce Hall current in Rashba spin-orbit coupled metal\cite{Edelstein} and graphene\cite{Oka}. Since circularly polarized light is described by ac electric field,  the Hall effect predicted in these works also requires the time-reversal symmetry breaking.

In this paper, we study transverse Josephson current in superconductor/normal metal/superconductor junctions where the normal metal has Rashba type spin-orbit coupling. 
It is shown that transverse current arises from the spin-orbit coupling in the normal metal under phase gradient, and the transverse current vanishes in the normal state.  In addition, this transverse Josephson effect is purely stationary and applied magnetic field is unnecessary to realize this effect, in contrast to the Hall effect in the normal state. 
We also discuss physical interpretation of this effect, comparing with the spin Hall effect where the transverse spin current is generated by the spin-orbit coupling in 2D electron system\cite{Murakami2,Sinova}. 

\begin{figure}[tbp]
\begin{center}
\scalebox{0.8}{
\includegraphics[width=8.50cm,clip]{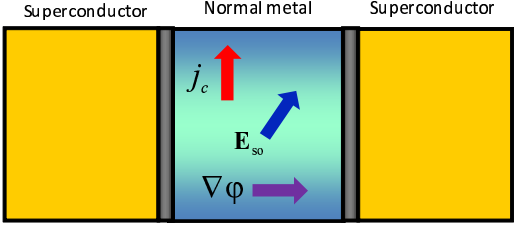}
}
\end{center}
\caption{ (Color online) Schematic picture of the model of superconductor/normal metal/superconductor
junctions where the normal metal has the Rashba type spin-orbit coupling characterized by the vector ${\bf{E}}_{so}$. The charge current ${\bf{j}}_{c}$ flows perpendicularly to the phase gradient ${\bm{\nabla}} \varphi$. }
\label{fig0}
\end{figure}

We consider a superconductor/normal metal/superconductor junction.
The Hamiltonian of the superconductor and the normal metal are given by  $H_S  = H_0  + H_\Delta $ and $H_N  = H_0  + H_{so}+H_\varphi + H_{so - \varphi } $, respectively. The superconductor and the normal metal are coupled via the tunneling Hamiltonian $H_T$ and the junction is described by a two barrier model. The total Hamiltonian of the system is thus given by $H_S+H_N+H_T$.
The $H_0$, $H_\Delta$ and  $H_{so}$ represent the kinetic energy, the superconducting order, and the Rashba type spin-orbit coupling, respectively:
\begin{eqnarray}
H_0  = \sum\limits_{\bf{k}} {\phi _{\bf{k}}^\dag  \xi \sigma _0  \otimes \tau _3 \phi _{\bf{k}}^{} } ,  \\ 
 H_\Delta   = \sum\limits_{\bf{k}} {\phi _{\bf{k}}^\dag  \Delta \sigma _0  \otimes \tau _1 \phi _{\bf{k}}^{} },  \\ 
H_{so}  =  - \sum\limits_{\bf{k}} {{\bf{E}}_{so}  \cdot \phi _{\bf{k}}^\dag  ({\bf{k}} \times {\bm{\sigma }}) \otimes \tau _3 \phi _{\bf{k}}^{} }   \label{hso} \end{eqnarray}
with $\xi  = \varepsilon _k  - \varepsilon _F  = \frac{{\hbar ^2 k^2 }}{{2m}} - \varepsilon _F$ and $\phi _{\bf{k}}^\dag   = (c_{{\bf{k}} \uparrow }^\dag  ,c_{{\bf{k}} \downarrow }^\dag  ,ic_{ - {\bf{k}} \downarrow }^{} , - ic_{ - {\bf{k}} \uparrow }^{} )$  where  $\sigma$ and $\tau$ are Pauli matrices in spin and Nambu spaces, respectively. $\varepsilon _F$, $\Delta$, and ${\bf{E}}_{so}$ are the Fermi energy, the gap function, and the vector pointing in the direction of the inversion symmetry breaking which characterizes the Rashba  type spin-orbit coupling, respectively. 
Note that we adopt the basis in Ref.\cite{Ivanov} such that singlet pairing is proportional to the unit matrix in spin space. 
We consider Josephson current induced by phase gradient.
The phase gradient along $j$ direction, $\nabla _j \varphi$, enters the Hamiltonian as follows
\begin{eqnarray}
H_\varphi   = \sum\limits_{\bf{k}} {\phi _{\bf{k}}^\dag  \frac{{\hbar ^2 }}{m}k_j \nabla _j \varphi \sigma _0  \otimes \tau_0 \phi _{\bf{k}}^{} } ,\\
H_{so - \varphi }  =  - \sum\limits_{\bf{k}} {{\bf{E}}_{so}  \cdot \phi _{\bf{k}}^\dag  \nabla _j \varphi ({\bf{e}}_j  \times {\bm{\sigma }}) \otimes \tau _0 \phi _{\bf{k}}^{} }
\end{eqnarray}
where $\varphi$ is half the phase of superconducting correlation and $\nabla _j \varphi$ is assumed to be spatially constant. We will treat $H_{so}$, $H_\varphi$, and $H_{so - \varphi } $ perturbatively. \cite{Hosono}
We schematically show the model in Fig. \ref{fig0}.

With the above Hamiltonians,
the charge current operator ($j_{c}$) in $i$-direction reads 
\begin{eqnarray}
j_{c,i}^{} =   - \frac{{e\hbar }}{m}k_i \sigma _0  \otimes \tau_0  - \delta _{ij} \frac{{e\hbar }}{m}\nabla _j \varphi \sigma _0  \otimes \tau _3  \nonumber \\
 + \frac{e}{\hbar } ({\bm{\sigma }} \times {\bf{E}}_{so} )_i \otimes \tau_0 \nonumber \\
 \equiv  j_{0,i}  + j_{\varphi ,i}  + j_{so,i} 
\end{eqnarray}
where $-e$ is the electron charge.

Before proceeding to the explicit calculation, let us discuss transverse current qualitatively based on the time-reversal symmetry. \cite{Murakami}
Now, consider the Ohm's law
\begin{equation}
{\bf{j}}_c  = \sigma  \cdot {\bf{E}}
\end{equation}
where ${\bf{j}}_c$, and ${\bf{E}}$ are, respectively,  the charge current and  the applied electric field. 
The charge current is time-reversal odd while the electric field is even under time-reversal. 
Since the Ohm's law relates quantities of different symmetries under time-reversal, the charge conductivity $\sigma$ breaks the time-reversal symmetry and describes the inevitable joule heating and dissipation.
A transverse current can flow under the applied electric field, which is hence dissipationless, but the time-reversal symmetry is compensated by the external magnetic field. Therefore, the Hall effect in the normal state occurs, inevitably accompanied by the time-reversal symmetry breaking. 

On the other hand, let us consider the London equation, i.e., the response equation of supercurrent, 
\begin{equation}
{\bf{j}}_c  =  - \frac{{e^2 }}{m}\rho  \cdot {\bf{A}}. \label{London}
\end{equation}
where  $\rho$ and ${\bf{A}}$ are, respectively, the superfluid density and  the vector potential. 
Since charge current and vector potential are time-reversal odd, $\rho$ describes the reversible and dissipationless flow of the supercurrent. Thus, transverse superconducting current can flow without breaking time-reversal symmetry.  
Similary, we see that since spin current is time-reversal even, superconducting spin current is absent without breaking time-reversal symmetry.\cite{Malshukov}

\begin{figure}[tbp]
\begin{center}
\scalebox{0.8}{
\includegraphics[width=10.50cm,clip]{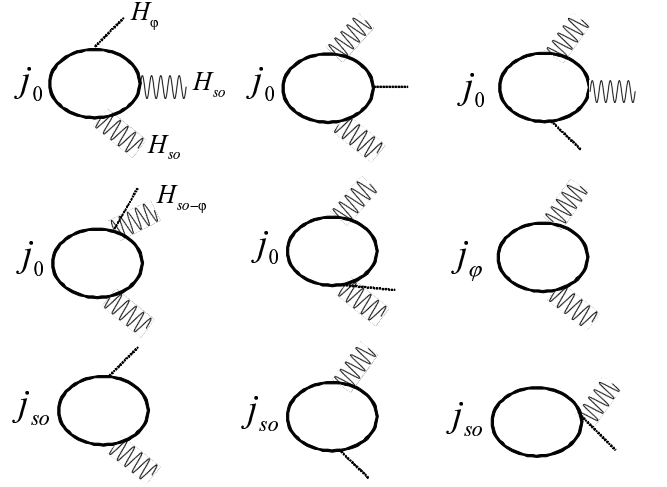}
}
\end{center}
\caption{ Diagrammatic representations of the current densities with second-order contributions of the spin-orbit coupling ${\bf{E}}_{so}$ and the first-order contributions of the phase gradient $\nabla \varphi$. The wavy lines denote the spin-orbit interaction and dotted lines represent the phase gradient. The wavy and dotted lines intersecting at one point correspond to $H_{so - \varphi }$. }
\label{fig1}
\end{figure}


Now, we calculate transverse Josephson current and give the analytical expression. We consider the unperturbed advanced Green's functions in the normal metal of the form 
\begin{eqnarray}
g_{{\bf{k}},\omega }^a  = g_{0,{\bf{k}},\omega }^a \sigma _0  \otimes \tau_0 + g_{3,{\bf{k}},\omega }^a \sigma _0  \otimes \tau _3  + f_{{\bf{k}},\omega }^a \sigma _0  \otimes \tau _1 \nonumber \\
 \equiv \sigma _0  \otimes g_{\tau ,{\bf{k}},\omega }^a \label{GF}
\end{eqnarray}
where $g_{0,{\bf{k}},\omega }^a$ and $g_{3,{\bf{k}},\omega }^a$ are normal Green's functions while $f_{{\bf{k}},\omega }^a $ is anomalous Green's function. The anomalous Green's function is in general nonzero in the normal metal due to the proximity effect. 
We perform perturbative calculation with respect to ${\bf{E}}_{so}$  and $\nabla _j \varphi$ up to a second and a first order, respectively. Diagrammatic representation of the transverse current is  shown in Fig. \ref{fig1}.
The transverse Josephson current can be expressed as  \cite{Haug}
\begin{eqnarray}
j_{c,i}  = \frac{{i\hbar ^2 e}}{{mV}}\sum\limits_{\bf{k}} { {\rm{Tr}}k_i \sigma _0  \otimes \tau_0 G_{{\bf{k}},{\bf{k}}}^ <  (t,t)} \nonumber \\ + \delta _{ij} \frac{{i\hbar ^2 e}}{{mV}}\nabla _j \varphi \sum\limits_{\bf{k}} {{\rm{Tr}}\sigma _0  \otimes \tau _3 G_{{\bf{k}},{\bf{k}}}^ <  (t,t)} \nonumber \\ 
 - \frac{{i e}}{{V}}\sum\limits_{\bf{k}} {{\rm{Tr}}({\bm{\sigma }} \times {\bf{E}}_{so} )_i  \otimes \tau_0  G_{{\bf{k}},{\bf{k}}}^ <  (t,t)} 
\end{eqnarray}
where $V$ is the total volume and ${\rm{Tr}}$ is taken over spin and Nambu spaces. $G_{{\bf{k}},{\bf{k}} }^ <  (t,t)$ is the lesser Green's function of the total Hamiltonian. 
Performing perturbation with respect to $H_{so}$, $H_\varphi$ and $H_{so-\varphi}$, we expand the lesser component using the advanced Green's functions by the Langreth theorem.\cite{Haug}
Noting that $g_{{\bf{k}},\omega }^ <   = f_\omega  \left[ {g_{{\bf{k}},\omega }^a  - (g_{{\bf{k}},\omega }^a )^\dag  } \right]$ with the lesser Green's function $g_{{\bf{k}},\omega }^ <$  and the Fermi distribution function $f_\omega$, and $\delta _{ij}  = \frac{{\partial k_i }}{{\partial k_j }}$, we can compute the transverse Josephson current.
Then, the leading  term of the transverse current ($i \ne j$) is given by the second order expansion with respect to ${\bf{E}}_{so}$ (the first order term vanishes), which results in the form 
\begin{widetext}
\begin{eqnarray}
 j_{c,i} \cong \frac{{2i\hbar ^4 e^2 }}{{Vm^2 }}\nabla _j \varphi \sum\limits_{{\bf{k}},\omega } {{\rm{Tr}_\tau}} k_i k_j ({\bf{E}}_{so}  \times {\bf{k}})^2  \left[ {(g_{\tau,{\bf{k}},\omega }^{} )^2 \tau _3 g_{\tau,{\bf{k}},\omega }^{} \tau _3 g_{\tau,{\bf{k}},\omega }^{}  + \tau _3 g_{\tau,{\bf{k}},\omega }^{} \tau _3 (g_{\tau,{\bf{k}},\omega }^{} )^2 \tau _3 g_{\tau,{\bf{k}},\omega }^{}  + g_{\tau,{\bf{k}},\omega }^{} \tau _3 g_{\tau,{\bf{k}},\omega }^{} \tau _3 (g_{\tau,{\bf{k}},\omega }^{} )^2 } \right]_{}^ <      \nonumber \\ 
  + \delta _{ij} \frac{{2i\hbar ^2 e^{} }}{{Vm}}\nabla _j \varphi \sum\limits_{{\bf{k}},\omega } {{\rm{Tr}_\tau}} \tau _3 ({\bf{E}}_{so}  \times {\bf{k}})^2 \left[ {g_{\tau,{\bf{k}},\omega }^{} \tau _3 g_{\tau,{\bf{k}},\omega }^{} \tau _3 g_{\tau,{\bf{k}},\omega }^{} } \right]_{}^ <   \nonumber  \\ 
 - \frac{{2i\hbar ^2 e^{} }}{{Vm}}\nabla _j \varphi {\bf{E}}_{so}^i {\bf{E}}_{so}^j \sum\limits_{{\bf{k}},\omega } {{\rm{Tr}_\tau}} k_j^2 \left[ {(g_{\tau,{\bf{k}},\omega }^{} )^2 \tau _3 g_{\tau,{\bf{k}},\omega }^{}  + g_{\tau,{\bf{k}},\omega }^{} \tau _3 (g_{\tau,{\bf{k}},\omega }^{} )^2 } \right]_{}^ <   \nonumber  \end{eqnarray}
\begin{eqnarray}
 = \frac{{256e^{} }}{{9V}}\nabla _j \varphi {\bf{E}}_{so}^i {\bf{E}}_{so}^j \sum\limits_{{\bf{k}},\omega } {f_\omega  \varepsilon _k {\mathop{\rm Im}\nolimits} \left[ {\varepsilon _k (f_{{\bf{k}},\omega }^a )^2 \left\{ {(g_{0,{\bf{k}},\omega }^a )^2  - (f_{{\bf{k}},\omega }^a )^2  + 5(g_{3,{\bf{k}},\omega }^a )^2 } \right\} + \frac{3}{2}g_{3,{\bf{k}},\omega }^a (f_{{\bf{k}},\omega }^a )^2 } \right]}  \label{jc}
\end{eqnarray}
\end{widetext}
where ${\rm{Tr}_\tau}$ means the trace in Nambu space (here we have already taken the trace in spin space). 
This is a general expression which is applicable to any Green's function of the form of Eq. (\ref{GF}). 
We see that in the absence of the superconductivity $f_{\bf{k}}^a  \to 0$, the transverse current vanishes $j_{c,i}  \to 0$. 
From Eq.(\ref{jc}),  we find that the direction of the transverse Josephson current is determined by the vector ${\bf{E}}_{so}$, which characterizes the spin-orbit coupling, as 
\begin{eqnarray}
j_{c,i}  \propto \left[ {{\bf{E}}_{so}^{}  \times ({\bf{E}}_{so}^{}  \times {\bm{\nabla}} \varphi )} \right]_i  \label{jcc}
\end{eqnarray}
for ${\bf{j}}_c  \bot {\bm{\nabla}} \varphi $. 
Longitudinal Josephson current can flow under phase gradient without spin-orbit coupling. Thus, in the leading order, it is proportional to the phase gradient and does not depend on spin-orbit coupling. Then, Eq.(\ref{jcc}) can be rewritten as 
\begin{eqnarray}
j_{c,i}  \propto \left[ {{\bf{E}}_{so}^{}  \times ({\bf{E}}_{so}^{}  \times {\bf{j}}_{c}^l )} \right]_i 
\end{eqnarray}
where ${\bf{j}}_{c}^l$ is the longitudinal Josephson current parallel to ${\bm{\nabla}} \varphi$. We also see that when the phase gradient is along $x$-axis, to obtain finite $j_{c,y} $ the  vector ${\bf{E}}_{so}$ should have both $x$ and $y$ components. 

When the normal metal is sufficiently thin and the interfaces between the normal metal and the superconductors are transparent, proximity effect becomes very strong such that the Green's functions in the normal metal have the same form as those in the bulk superconductor:
\begin{eqnarray}
g_{{\bf{k}},\omega }^a   = \frac{{(\omega  - i\gamma)\sigma _0  \otimes \tau _0  + \xi \sigma _0  \otimes \tau _3  + \Delta \sigma _0  \otimes \tau _1 }}{{(\omega  - i\gamma )^2  - \xi ^2  - \Delta ^2 }}.
\end{eqnarray}
The transverse current is then given by 
\begin{eqnarray}
 j_{c,i} & & \cong  \frac{{64\pi e^{} \nu }}{{9V}}\nabla _j \varphi {\bf{E}}_{so}^i {\bf{E}}_{so}^j \sum\limits_\omega  {f_\omega  {\mathop{\rm Re}\nolimits} \left[ {\frac{{\Delta ^2 }}{{\left[ {(\omega  - i\gamma )^2  - \Delta ^2 } \right]^{3/2} }}} \right]}  \nonumber \\ 
 & & = \frac{{32e^{} \nu }}{{9V\hbar }}\nabla _j \varphi {\bf{E}}_{so}^i {\bf{E}}_{so}^j \left( {\frac{{\gamma }}{{\sqrt {\gamma ^2  + \Delta ^2 } }} - 1} \right) \label{jcb}
\end{eqnarray}
at zero temperature where $\gamma$ is the inelastic scattering rate by impurities and $\nu$ is the density of states at the Fermi energy.

\begin{figure}[tbp]
\begin{center}
\scalebox{0.8}{
\includegraphics[width=10.60cm,clip]{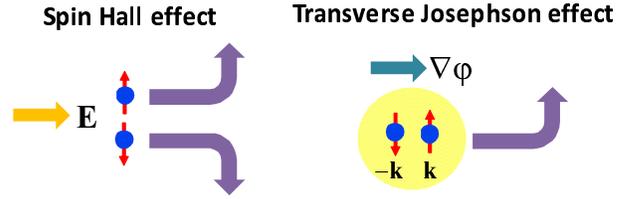}}
\end{center}
\caption{(Color online) Physical picture of the spin Hall effect (left) and the transverse Josephson effect by Cooper pair (right). }
\label{fig2}
\end{figure}

Let us discuss the mechanism of the transverse Josephson effect predicted in this paper, comparing with the spin Hall effect\cite{Murakami2,Sinova}. 
Consider the spin-orbit coupling Eq.(\ref{hso}) as a spin dependent potential. The  spin Hall effect occurs since electron with spin-up and that with spin-down feel potential with the opposite sign and hence are scattered in the opposite direction as shown in Fig. \ref{fig2}. On the other hand, the Josephson current is carried by Cooper pair which consists of electron pair with opposite momentum and spin, $({\bf{k}}, \uparrow )$ and $( - {\bf{k}}, \downarrow )$. As seen from Eq.(\ref{hso}), the Rashba type spin-orbit coupling is invariant under the sign change of both momentum and spin (since this term is time-reversal even). Thus, the two electrons which constitute the Cooper pair feel exactly the same potential and hence move in the same direction, resulting in the net transverse current.

As seen from the London equation Eq.(\ref{London}), since both charge current and vector potential are  odd under spatial inversion, the inversion symmetry breaking is unnecessary to obtain the charge current. In fact, the coefficient of the phase gradient is given by even (second) order with respect to the Rashba type spin-orbit coupling (see Eq.(\ref{jc})),  and hence does not break the inversion symmetry.
This implies that when the system respects the inversion symmetry, the transverse Josephson effect predicted in this paper would also emerge due to spin-orbit coupling. 
Further investigation of the transverse Josephson effect based on such model as the Luttinger Hamiltonian\cite{Luttinger}, the spin-orbit coupling of which preserves the inversion symmetry, will be an interesting future work.

The present formalism assumes the translational symmetry of the system and hence the length of the normal metal does not appear in the results. To investigate the effect of the length of the normal metal, investigation based on, for example, the Usadel equation including spin-orbit coupling is necessary. \cite{Malshukov} 

To test the transverse Josephson effect predicted in this paper experimentally, one may use BiTeI, a recently discovered 3D Rashba system\cite{Ishizaka,Bahramy}, as a spin-orbit coupled normal metal. 
For this material, we have $\nu \sim 5 \times 10^{-2}$ states/eV/unit cell, the lattice constant $\sim 5$ \AA, $k_F {{E}}_{so} \sim$ 100 meV, and the Fermi wavevector $k_F \sim$ 0.1 \AA$^{-1}$. \cite{Ishizaka,Bahramy}
Taking $\nabla _j \varphi \sim (100  \rm{nm})^{-1}$, $\gamma \sim$ 10 meV and $\Delta \sim$ 1 meV, we estimate the current in Eq.(\ref{jcb}) as $\left| {j_{c,i}^{} } \right| \sim 2 \times 10^{-4}$ A/cm$^2$. 
Our prediction could be confirmed by Josephson junctions with this material in four-terminal geometry by injecting longitudinal current. \cite{comment} 
Since transverse Josephson current gives a phase shift along the transverse direction, one can detect the predicted effect by using superconducting quantum interference device attached to the normal metal transversely. The transverse Josephson effect is reflected as a shift in the interference pattern.

In summary,
 we have investigated transverse Josephson current in superconductor/normal metal/superconductor junctions where the normal metal has Rashba type spin-orbit coupling.  It has been shown that transverse current arises from the spin-orbit coupling in the normal metal. This effect is specific to superconducting current and the transverse current vanishes in the normal state.  In addition, this transverse Josephson effect is purely stationary and applied magnetic field is unnecessary to realize this effect, in contrast to the Hall effect in the normal state. 
We have also presented physical interpretation of this effect, comparing with the spin Hall effect.


The author thanks S. Murakami for helpful discussion.
This work was supported by Grant-in-Aid for Young Scientists (B) (No. 23740236) and the "Topological Quantum Phenomena" (No. 23103505) Grant-in Aid for Scientific Research on Innovative Areas from the Ministry of Education, Culture, Sports, Science and Technology (MEXT) of Japan.

\textit{Appendix.}
Here, we will derive Eq.(6). 
The velocity operator in $i$-direction, $v_i$, is given by the derivative of the Hamiltonian with respect to the momentum: 
\begin{eqnarray}
v_i  = \frac{{\partial H_N }}{{\hbar \partial k_i }} 
= \frac{\hbar }{m}k_i \sigma _0  \otimes \tau _3  + \delta _{ij} \frac{\hbar }{m}\nabla _j \varphi \sigma _0  \otimes \tau _0   \nonumber \\
- \frac{1}{\hbar } ({\bm{\sigma }} \times {\bf{E}}_{so} )_i  \otimes \tau _3 .
\end{eqnarray}
One obtains the current operator by multiplying the velocity operator by the charge operator $- e\sigma _0  \otimes \tau _3 $ :
\begin{eqnarray}
j_{c,i}^{}  =  - e\sigma _0  \otimes \tau _3 v_i  
=  - \frac{{e\hbar }}{m}k_i \sigma _0  \otimes \tau _0  \nonumber \\ 
- \delta _{ij} \frac{{e\hbar }}{m}\nabla _j \varphi \sigma _0  \otimes \tau _3   + \frac{e}{\hbar } ({\bm{\sigma }} \times {\bf{E}}_{so} )_i  \otimes \tau _0 .
\end{eqnarray}
Note that the charge is opposite for electron and hole. 


\begin{thebibliography}{99}
\bibitem{comment} Since the magnitude of the spin-orbit coupling of this material is large, the perturbative calculation with respect to the spin-orbit coupling may not be a good approximation. However, the qualitative result, the appearance of the transverse current, should remain.
\end{thebibliography}
\end{document}